\documentclass[twocolumn]{raa}
\usepackage{graphicx,times}
\usepackage{natbib}
\usepackage{amssymb,amsmath}
\bibpunct{(}{)}{;}{a}{}{,}

\usepackage[a4paper=true,driverfallback=dvipdfm,pagebackref=true]{hyperref}

\hypersetup{colorlinks = true, linkcolor = green, anchorcolor = red, citecolor = blue, filecolor = red, pagecolor = red, urlcolor = red}

\mathchardef\mhyphen="2D
\defcitealias{Henriques2015}{H15}

\begin{document}
 \title{Halo Spin Dependence on Environment for HI-bearing galaxies}

 \volnopage{ {\bf 20XX} Vol.\ {\bf X} No. {\bf XX}, 000--000}
   \setcounter{page}{1}

\author{Zichen Hua\inst{1,2}, Yu Rong\inst{1,2}\thanks{corresponding author}, Huijie Hu\inst{3}
      }


   \institute{ Department of Astronomy, University of Science and Technology of China, Hefei, Anhui 230026, China; {\it rongyua@ustc.edu.cn}\\
         \and
             School of Astronomy and Space Sciences, University of Science and Technology of China, Hefei 230026, Anhui, China\\
           \and
                  University of Chinese Academy of Sciences, Beijing 100049, China\\
 \vs \no
   {\small Received 20XX Month Day; accepted 20XX Month Day}
}

\abstract{Leveraging the semi-analytic method, we compute halo spins for a substantial sample of HI-bearing galaxies observed in the Arecibo Legacy Fast Alfa Survey. Our statistical analysis reveals a correlation between halo spin and environment, although the trend is subtle. On average, galaxies exhibit a decreasing halo spin tendency in denser environments. This observation contrasts with previous results from $N$-body simulations in the Lambda cold dark matter framework. The discrepancy may be attributed to environmental gas stripping, leading to an underestimation of halo spins in galaxies in denser environments, or to baryonic processes that significantly alter the original dark matter halo spins, deviating from previous $N$-body simulation findings.
}

   \authorrunning{Hua et al.}            
   \titlerunning{halo spin dependence on environment}  
   \maketitle
%
\section{Introduction}           
\label{sec:1}

In the context of Lambda cold dark matter cosmology, it is typically assumed that the distributions of galaxies/baryons reflect the distributions of dark matter. Gas is accreted into self-bound, virialized dark matter halos, where it undergoes cooling primarily through radiation and condenses into the central regions of the halos, eventually transforming into stars to give rise to galaxies. The dark matter halo plays a pivotal role in galaxy formation by providing the gravitational potential for gas condensation, serving as a stage for shock-heating gas, and contributing to galaxy rotation through its spin.

Halo spin has significant implications for galaxy evolution and morphology. Hydrodynamical simulations (e.g., \citealt{Kim13}) and semi-analytic galaxy formation models (e.g., \citealt{Mo98}) indicate that halo spin strongly influences the size and density of baryonic matter distribution, especially in massive late-type galaxies. While the role of halo spin in low-mass galaxies remains debated, studies on specific dwarf galaxies, such as ultra-diffuse galaxies (\citealt{Rong17a,Amorisco16,Liao19,Benavides23}), suggest that halo spin may significantly impact the distribution of baryonic matter in dwarf galaxies.

Halo spin is believed to arise from tidal torques exerted by large-scale structures, resulting from gravitational interactions with neighboring structures (e.g., \citealt{Peebles69,White84}), or through mergers (e.g., \citealt{Gardner01,Vitvitska02,Hetznecker06,Maller02}). As halos approach the turnaround stage between linear and non-linear phases and eventually reach virialization, the influence of tidal torque diminishes. During this phase, the flow field surrounding halos exhibits non-zero vorticity, crucial in determining halo angular momentum and leading to alignment between halo spin and vorticity (e.g., \citealt{Liberskind13}). Notably, halo spin is influenced by the surrounding environment \citep{Hahn07a,Hahn07b,Wang17,Wang18}. Simulation studies indicate that halos tend to spin faster in stronger tidal fields, with a more pronounced effect observed in more massive halos (\citealt{Wang11}). Understanding dark matter halo spin is essential for unraveling the formation and evolution of galaxies in the universe.

However, determining halo spin is a challenging task observationally. Typically, it is estimated by analyzing the velocity fields of stellar or gas components (e.g., \citealt{Wang20,Rong18}). Presently, only a limited number of high surface brightness galaxies have spatially resolved data with high signal-to-noise ratios suitable for halo spin calculations. Alternatively, broad HI surveys conducted with single-dish telescopes, such as the comprehensive Arecibo Legacy Fast Alfa Survey  \citep[ALFALFA;][]{Giovanelli05,Haynes18} and undergoing FAST All Sky HI survey \citep[FASHI;][]{Zhu23}, offer a valuable opportunity to acquire HI spectra from numerous galaxies. These surveys provide essential dynamical information on galaxies, allowing for the estimation of spin parameters across a large galaxy sample and enabling comparisons of rotational speeds in diverse environments.

In this investigation, we employ a semi-analytic approach to estimate halo spin for each HI-bearing galaxy cataloged in ALFALFA. Section~\ref{sec:2} presents the sample data and outlines the methodology for estimating halo spin. Section~\ref{sec:3} offers a statistical comparison of galaxy halo spins in different environments and discusses the outcomes. Our findings are summarized in section~\ref{sec:4}.

\section{Data}
\label{sec:2}

\subsection{Sample}

The galaxy sample is drawn from ALFALFA, an extensive extragalactic HI survey spanning approximately 6,600 deg$^2$ at high Galactic latitudes. The ALFALFA collaboration has released a fully comprehensive catalog \citep[$\alpha.$100;][]{Haynes18} comprising around 31,500 sources with radial velocities below 18,000 km s$^{-1}$. This catalog includes various properties for each source, such as the signal-to-noise ratio (SNR) of the HI spectrum, cosmological distance, 50\% peak width of the HI line ($W_{50}$) corrected for instrumental broadening, and the HI mass ($M_{\rm{HI}}$), among others. For detailed definitions of these parameters (and their uncertainties) and the estimation method, we direct readers to \cite{Haynes18}.

\subsection{Stellar mass}

ALFALFA galaxies have been matched with SDSS data \citep{Alam15}. Previous studies by\cite{Durbala20} have estimated the stellar masses $M_{\star}$ of ALFALFA galaxies with optical counterparts using three methods: UV-optical-infrared SED fitting, SDSS $g-i$ color, and infrared $W_2$ magnitude. This study prioritizes the stellar mass derived from SED fitting. In cases where UV or infrared data are unavailable for SED fitting, leading to an inability to estimate the stellar mass, the stellar mass based on $g-i$ color is utilized. Any discrepancies in stellar mass obtained from these methods are deemed insignificant.

It is important to note that ``dark galaxies'', a subset of ALFALFA galaxies without optical counterparts or displaying faint optical signatures, have been excluded from this study. These ``dark galaxies'',  as described by \cite{Disney76} and \cite{Janowiecki15}, are known to be prone to tidal interactions \citep{Roman21,Duc08}, making them non-equilibrium systems and leading to inaccurate estimates of rotation velocities. 

\subsection{Rotation velocity}

The rotation velocity is calculated as $V_{\rm{rot}}=W_{50}/2/\sin\phi$, where $\phi$ represents the inclination of the HI disk. In cases where resolved HI data is unavailable, this study utilizes the optical apparent axis ratio $b/a$, as provided by \cite{Durbala20}, to estimate the HI disk inclination $\phi$. The calculation is done using $\sin\phi=\sqrt{(1-(b/a)^2)/(1-q_0^2)}$ (if $b/a \leq q_0$, we set $\phi=90^{\circ}$), with the intrinsic thickness $q_0\sim 0.2$ \citep{Tully09,Giovanelli97,Li21} for massive galaxies, and $q_0\sim 0.4$ (\citealt{Rong24}) for low-mass galaxies with $M_{\star}<10^{9.5}\ M_{\odot}$.

To enhance the accuracy of rotation velocity estimation, galaxies with inclinations $\phi<50^{\circ}$ are excluded. Additionally, galaxies with low HI signal-to-noise ratios (SNR$<20$) are also excluded due to significant uncertainties in velocity estimation.

As a result, a sample of approximately $7,600$ galaxies is obtained. The stellar masses of our sample range from around $10^7$ to $10^{11}$ $\rm M_{\odot}$, as illustrated in panel~a of Fig.~\ref{fig1}. 

\subsection{Halo spin}

Theoretical considerations, assuming a galaxy's dark matter halo follows the isothermal sphere model and neglecting the gravitational impact of baryonic matter, allow us to express the galaxy's halo spin as (\citealt{Hernandez07}):
\begin{equation}
\lambda_{\rm{h}}\simeq 21.8 \frac{R_{\rm{HI,d}}/{\rm kpc}}{(V_{\rm{rot}}/{\rm km s^{-1}})^{3/2}},
	\label{sam_HI} \end{equation}
Here, $V_{\rm{rot}}$ denotes the halo's rotation velocity.
The scale length of the HI disk, $R_{\rm{HI,d}}$, can be determined by assuming a relatively thin gas disk in centrifugal balance \citep{Mo98}, characterized by an exponential surface density profile:
\begin{equation}
	\Sigma_{\rm{HI}}(R)=\Sigma_{{\rm{HI}},0} {\rm{exp}}(-R/R_{{\rm{HI,d}}}),
\end{equation}
where $\Sigma_{{\rm{HI}},0}$ is the central surface density of the HI disk. The total HI mass $M_{\rm{HI}}$ is linked to the scale length as
\begin{equation} M_{\rm{HI}} = 2 \pi \Sigma_{{\rm{HI}},0} R_{{\rm{HI,d}}}^2 \label{HIeq_mass}. \end{equation} 
Additionally, we introduce the HI radius $r_{\rm{HI}}$, defined as the radius at which the HI surface density reaches $1\ \rm M_{\odot}\rm{pc^{-2}}$. The estimation of $r_{\rm{HI}}$ is guided by the observed correlation between $r_{\rm{HI}}$ and HI mass $M_{\rm{HI}}$, as indicated by empirical studies: $\log r_{\rm{HI}}=0.51\log M_{\rm{HI}}-3.59$ \citep{Wang16,Gault21}. Therefore, at $r_{\rm{HI}}$, we have 
\begin{equation} \Sigma_{{\rm{HI}},0} {\rm{exp}}(-r_{\rm{HI}}/R_{{\rm{HI,d}}})=1\ \rm  M_{\odot}\rm{pc^{-2}}. \label{HIeq_3} \end{equation} 
By utilizing equations~(\ref{HIeq_mass}) and (\ref{HIeq_3}), we can compute the value of $R_{{\rm{HI,d}}}$ for each galaxy in our sample, thereby enabling the estimation of the halo spin.

\subsection{Environment}

To assess the environmental context of each galaxy in our sample, we utilize the galaxy group and cluster catalog developed by \cite{Saulder16}. This catalog, derived from the SDSS DR12 \citep{Alam15} and 2MASS Redshift Survey \citep{Huchra12}, employs the friends-of-friends group finder algorithm. Notably, the work by \cite{Saulder16} meticulously addresses various observational biases, including the Malmquist bias and the `Fingers of God' effect. 

For environmental classification of galaxies, we adopt the criteria outlined by \cite{Rong24a}. Specifically, galaxies are deemed isolated if they reside beyond three times the virial radius of any galaxy group or cluster. Conversely, galaxies failing to meet this criterion are classified as nonisolated.

\begin{figure}
   \centering
   \includegraphics[width=\columnwidth, angle=0]{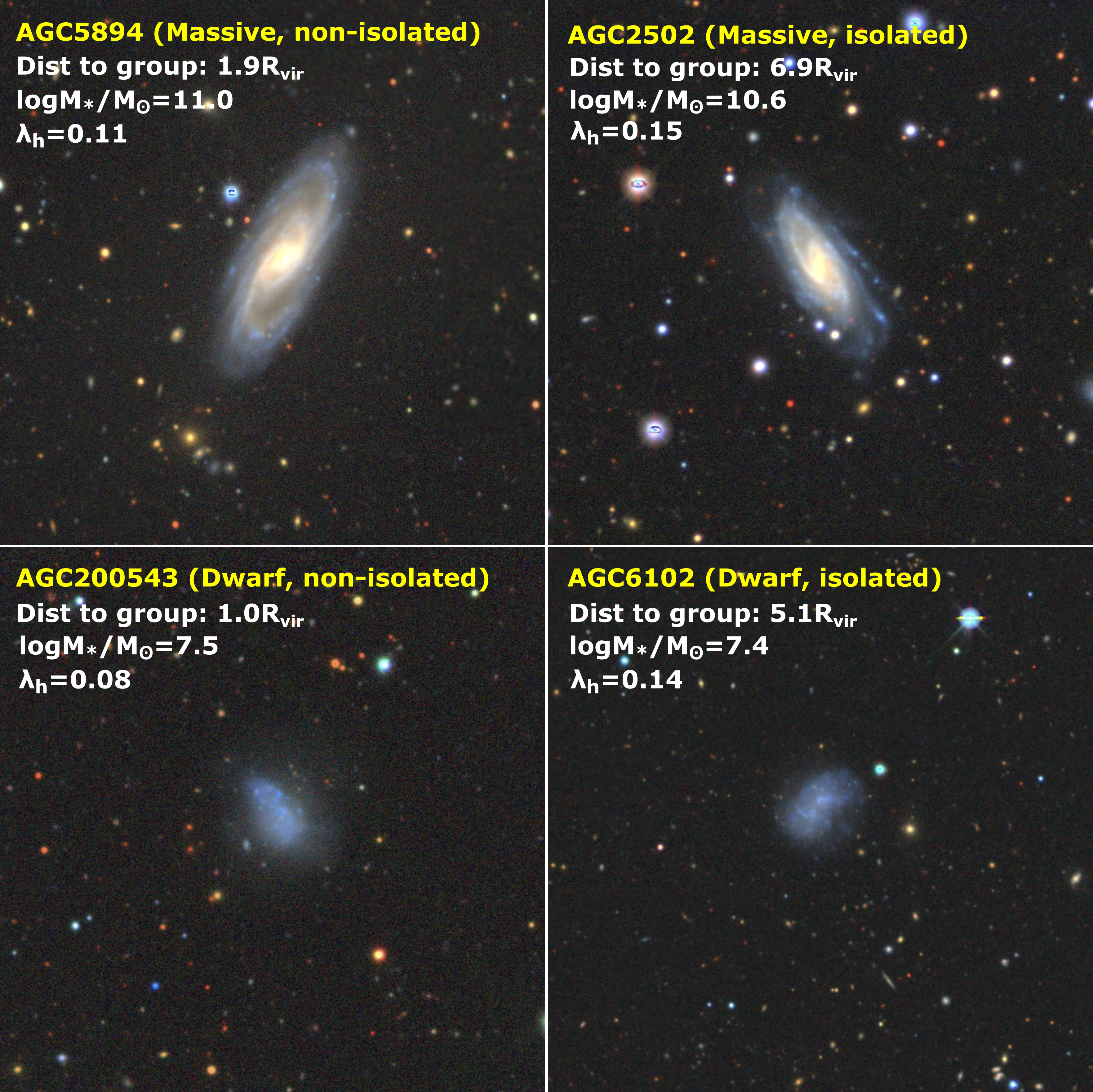}
   \caption{The DESI images depict two example massive galaxies (upper panels) and two example low-mass galaxies (lower panels) situated within three times the virial radii of galaxy groups (left panels) and fields (right panels). Each panel also displays the distance to the nearest galaxy group (normalized by the group's virial radius), stellar mass, and halo spin of the respective galaxy.}
   \label{pic}
   \end{figure}

In Fig.~\ref{pic}, we show the optical images of four example sample galaxies located in the different environments (isolated and non-isolated) and with different stellar masses (low-mass and high-mass).


\section{Results and discussion}
\label{sec:3}

\begin{figure*}
   \centering
   \includegraphics[width=0.9\textwidth, angle=0]{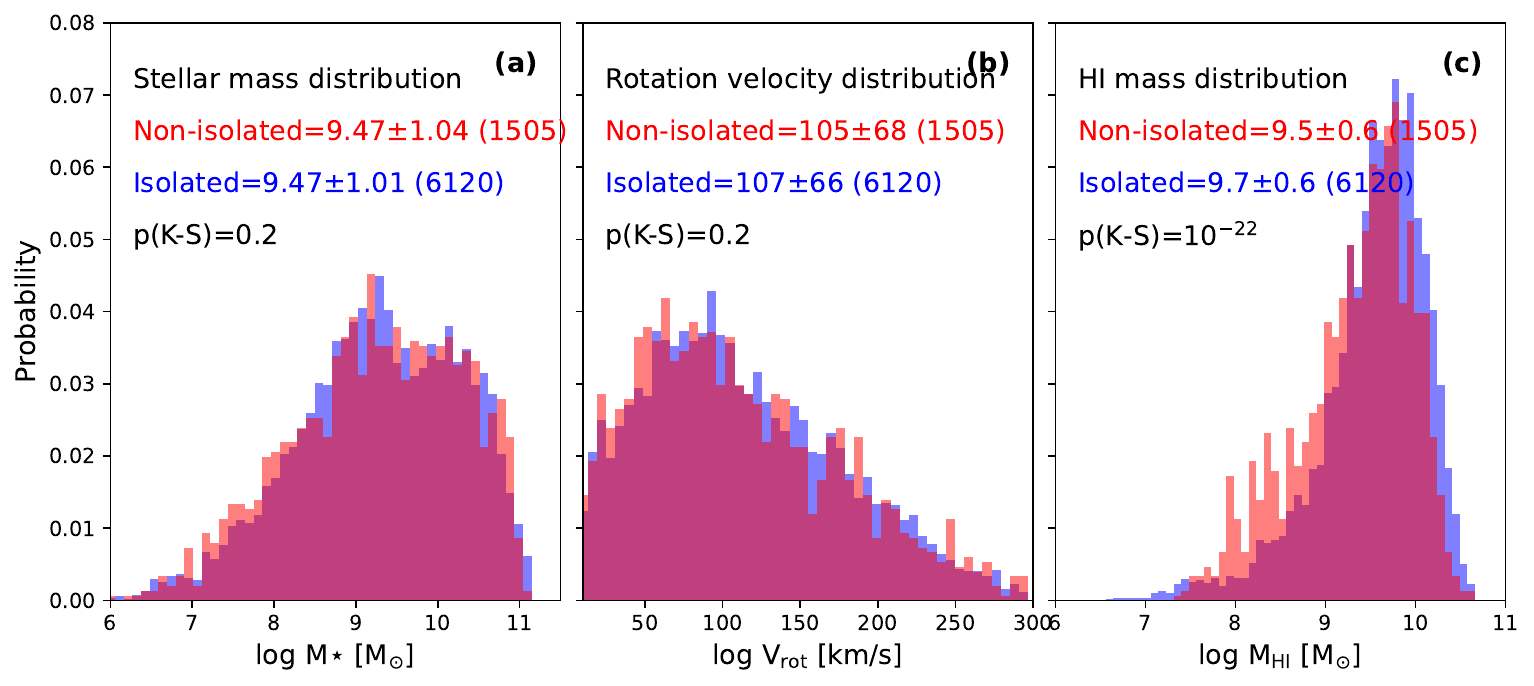}
   \caption{The distributions of stellar mass (panel~a), rotation velocity (panel~b), and HI mass (panel~c) for the isolated (blue) and non-isolated (red) galaxies. The median values and corresponding sample sizes (in brackets) are shown, along with the K-S test $p$-values of comparing the isolated and non-isolated galaxy distributions.}
   \label{fig1}
   \end{figure*}

\begin{figure*}
   \centering
   \includegraphics[width=\textwidth, angle=0]{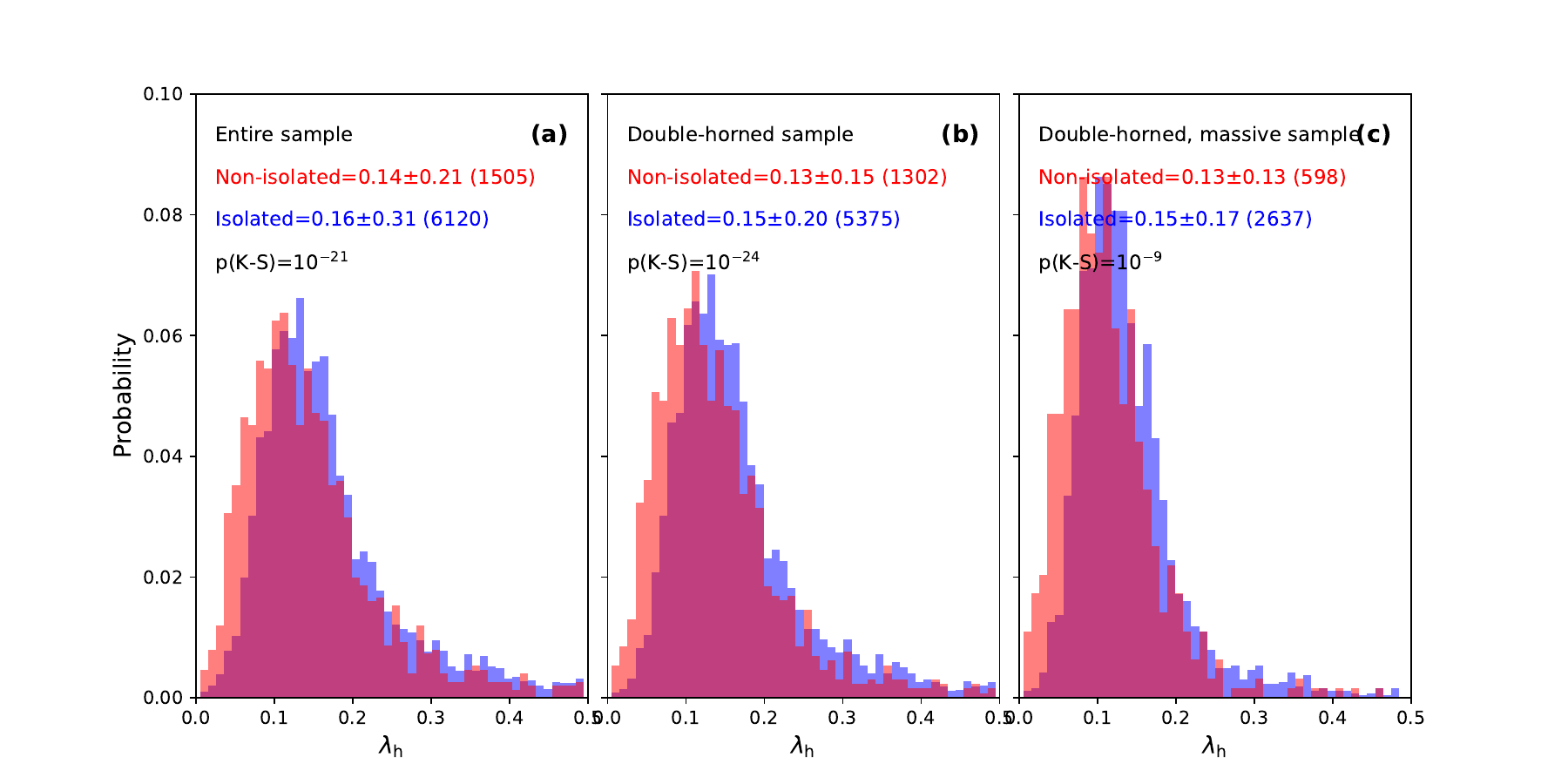}
   \caption{The comparison between the halo spin distributions of the isolated (blue) and non-isolated (red) HI-bearing galaxies. Panels~a, b, and c display the entire galaxy sample, sample with double-horned HI spectra, and massive galaxy sample, respectively. The median value and $1\sigma$ uncertainty for the corresponding distribution are shown, along with the sample size in the bracket. The small $p$-values from K-S tests of comparing the isolated and non-isolated halo spin distributions indicate the spin distribution difference.}
   \label{compare}
   \end{figure*}

In panel~a of Fig.~\ref{compare}, we present a comparison of the halo spin distributions between isolated galaxies (blue histogram) and non-isolated galaxies (red histogram).
Our analysis reveals a slight decrease in halo spins for galaxies located in denser environments. The results of the Kolmogorov-Smirnov (K-S) test further validate distinct halo spin distributions between the two subgroups across various environmental conditions. Specifically, the median halo spin of non-isolated galaxies is approximately 0.02 lower than that of isolated galaxies. This spin discrepancy may stem from environmental gas stripping, as the HI masses of non-isolated galaxies are statistically lower than those of their isolated counterparts (panel~c of Fig.~\ref{fig1}), or potentially due to baryonic processes significantly altering the original dark matter halo spins, deviating from previous findings of $N$-body simulations \citep[e.g.,][]{Wang11,Hahn07b}.

Initially, it is crucial to acknowledge that halo spin is closely linked to galaxy mass. Therefore, if the galaxy masses differ between subsamples, $\lambda_{\rm h}$ may also exhibit discrepancies. However, as depicted in panels~a and b of Fig.~\ref{fig1}, the stellar masses and rotation velocities (representing halo mass; \citealt{Mo98}) of the two subsamples are notably similar. This similarity is evidenced by the high $p$-values from K-S tests and the comparable median values of the distributions. Consequently, the observed spin variation with respect to environment cannot be solely attributed to differences in stellar mass or halo mass between the subsamples.

Secondly, it is important to recognize that certain HI-bearing galaxies may be characterized by a dominance of velocity dispersion over regular rotation. These galaxies, identified by their HI line profiles exhibiting a `single-horned' shape \citep{ElBadry18}, pose challenges in accurately estimating rotation velocities and, consequently, halo spins. To distinguish between single-horned and double-horned HI spectra, we employ the kurtosis parameter $k_4$, following the methodology outlined in \cite{Hua24}. A spectrum is typically classified as single-horned if $k_4>-1.0$.

Within our sample, approximately 12\% exhibit single-horned HI line profiles with $k_4>-1.0$. We focus on the robust subsets of isolated galaxies with double-horned HI profiles, excluding potentially dispersion-dominated galaxies. The results, as depicted in panel~b of Fig.~\ref{compare}, align with those of the complete dataset.

Furthermore, it is worth noting the presence of small misalignments ${\rm d}\phi$ between optical and HI inclinations observed in numerous galaxies \citep{Hunter12,Oh15} and simulations \citep{Nelson18,Nelson19,Vogelsberger14}. However, the majority ($\gtrsim 70\%$) of galaxies exhibit ${\rm d}\phi<20^{\circ}$. In statistical terms, this misalignment would primarily increase the scatter within the spin distribution of each galaxy subsample, rather than bias the spin estimation (e.g., elevating or reducing spins across the entire sample). Hence, the observed spin variation with respect to environment cannot be attributed to potential inclination misalignments.

Finally, given the inclusion of numerous low-mass galaxies in our sample, where equation~(\ref{sam_HI}) may not be directly applicable \citep{Yang23}, we investigate the halo spin dependence for massive galaxies with stellar masses $M_{\star}>10^{9.5}\ M_{\odot}$, as depicted in panel~c of Fig.~\ref{compare}. The results are consistent with those of the overall sample.

\section{Summary}
\label{sec:4}

Utilizing the semi-analytic approach, we calculate the halo spins for a substantial sample of HI-bearing galaxies observed in ALFALFA. Our statistical analysis reveals a correlation between halo spin and environment, demonstrating that galaxies in denser environments tend to have lower halo spins on average. This trend could be linked to environmental gas stripping, potentially causing an underestimation of HI masses and consequent halo spins in non-isolated galaxies. Alternatively, baryonic processes may significantly modify the initial dark matter halo spins, deviating from previous results of $N$-body simulations. Further exploration with spatially resolved data is warranted.


\begin{acknowledgements}

Y.R. acknowledges supports from the CAS Pioneer Hundred Talents Program (Category B), and the NSFC grant 12273037, as well as the USTC Research Funds of the Double First-Class Initiative.
\end{acknowledgements}
  

\end{document}